\documentclass[aps,twocolumn,nofootinbib,showpacs,prd,aps,10pt]{revtex4}
\usepackage[dvips]{graphicx}
\usepackage[english]{babel}
\usepackage[T1]{fontenc}
\usepackage{mathrsfs}
\usepackage[tbtags]{amsmath}
\usepackage{amssymb}
\usepackage{amsxtra}
\usepackage{amsopn}
\usepackage{latexsym}
\usepackage[mathcal]{eucal}
\usepackage{mathtools}

\newcommand{\BE}{\begin{equation}}
\newcommand{\EE}{\end{equation}}
\newcommand{\BA}{\begin{align}}
\newcommand{\EA}{\end{align}}
\newcommand{\Tr}{\mathrm Tr}
\newcommand{\nn}{\nonumber}

\newcommand{\kkd}{ \frac{{\rm d}^dk}{(2\pi)^d}}

\begin{document}

\title{Analytical study of Yang-Mills theory in the infrared from first principles}

\author{Fabio Siringo}

\affiliation{Dipartimento di Fisica e Astronomia 
dell'Universit\`a di Catania,\\ 
INFN Sezione di Catania,
Via S.Sofia 64, I-95123 Catania, Italy}

\date{\today}

\begin{abstract}

Pure Yang-Mills SU(N) theory is studied in the Landau gauge and four dimensional space.
While leaving the original Lagrangian unmodified, a double perturbative expansion is 
devised, based on a {\it massive} free-particle propagator. In dimensional
regularization, all diverging mass terms cancel exactly in the double 
expansion, without the need to include mass counterterms that would spoil the symmetry 
of the Lagrangian. No free parameters are included that were not in
the original theory, yielding a fully analytical approach from first principles.
The expansion is safe in the infrared and is equivalent to the standard perturbation theory in the UV. 
At one-loop, explicit analytical expressions are given for the propagators and the running coupling and 
are found in excellent agreement with the data of lattice simulations. 
A universal scaling property is predicted for the inverse propagators and shown to be satisfied by
the lattice data. Higher loops are found to be negligible in the infrared below $300$ MeV where
the coupling becomes small and the one-loop approximation is under full control.
\end{abstract}




\maketitle

\maketitle

\section{introduction}

In modern textbooks on QCD, the infrared domain is usually called {\it non-perturbative}
just because standard perturbation theory breaks down at 
the low-energy scale $\Lambda_{QCD}\approx$ 200 MeV.
While the high energy behaviour of the theory is under control and an analytical study of
non-Abelian gauge theories is usually achieved by a loop expansion in the UV, no analytical 
first-principle description of the infrared can be found in books where the subject is usually
discussed by phenomenological models that rely on numerical lattice simulations.

In the last years, important progresses have been achieved by non-perturbative approaches
based on Schwinger-Dyson equations (SDE) 
\cite{aguilar8,aguilar10,aguilar14,aguilar14b,papa15,papa15b,huber14,huber15g,huber15b}, 
variational methods\cite{reinhardt04,reinhardt05,reinhardt08,reinhardt14,sigma,
sigma2,gep2,varqed,varqcd,genself,ptqcd}, Gribov copies\cite{dudal08,sorella15,dudal15} 
and by simulating
larger and larger lattices\cite{bogolubsky,twoloop,dudal,binosi15,burgio15} of course.
While we still miss a full analytical description, the numerical solution of truncated
sets of SDE integral equations together with the {\it measures} that come from the lattice
yield a more clear picture of the infrared behaviour of QCD and Yang-Mills theory.

It is now widely believed that in the Landau gauge the gluon propagator is finite and
an effective coupling can be defined that is infrared safe and relatively small.
As discussed by Cornwall\cite{cornwall} in 1982, the gluon may acquire a dynamical mass
in the infrared without breaking the gauge invariance of the theory. The effect cannot
be described by the standard perturbation theory at any finite order because of gauge 
invariance that makes the polarization transverse and prohibits any shift of the pole in the
gluon propagator. That is one of the reasons why the standard perturbation theory cannot
predict the correct phenomenology in the infrared.

An other reason is the occurrence of a Landau pole in the running of the coupling that makes
evident the failure of the perturbative expansion below $\Lambda_{QCD}$.
However, in the Landau gauge the ghost-gluon vertex function can be shown to be finite\cite{taylor}
and a running coupling can be defined by the product of two-point correlators. Being massive,
the gluon propagator is finite and its dressing function vanishes in the infrared yielding a finite
running coupling that reaches a maximum and decreases in the low-energy limit\cite{bogolubsky}.
On the other hand, if the Landau pole is an artifact of the perturbative expansion, the relatively
small value of the real coupling suggests that we could manage to set up a different perturbative scheme
in the infrared. Actually, in order to make physical sense, a perturbative expansion requires that
the lowest order term should approximately describe the exact result. While that condition is fulfilled by
the standard perturbation theory in the UV, where the propagator is not massive, the dynamical
mass of the gluon makes the free propagator unsuitable for describing the low energy limit.
Thus we would expect that, by a change of the expansion point, a perturbative approach to QCD in the
infrared could be viable. 

There is some evidence that inclusion of a mass by hand 
in the Lagrangian gives a phenomenolgical model that describes very well the lattice data 
in the infrared at one loop\cite{tissier10,tissier11,tissier14}.
However that model could be hardly justified by first principles because of the mass that breaks the
gauge invariance of the Lagrangian. Even in a fixed gauge, BRST symmetry is broken and a mass counterterm
must be included for renormalizing the theory, thus introducing spurious free parameters in the model.

A change of the expansion point can be achieved by first principles without changing the original
Lagrangian. Variational calculations have been proposed\cite{gep2,varqed,varqcd,genself} 
where the zeroth order propagator is a trial
unknown function to be determined by some set of stationary conditions. The added propagator is
subtracted in the interaction, leaving the total action unchanged. The idea is not new and goes
back to the works on the Gaussian effective potential\cite{stevenson,var,light,bubble,ibanez,
su2,LR,HT,superc1,superc2,AF,stancu2,stancu}
where an unknown mass parameter was inserted
in the zeroth order propagator and subtracted from the interaction, yielding a pure variational
approximation with the mass that acts as a variational parameter.
Some recent variational calculations on Yang-Mills theory\cite{varqcd,genself} have shown that,
provided that we change the expansion point, a fair agreement with the lattice data can be
achieved without too much numerical effort. Thus we expect that it is not very important the actual
choice of the zeroth order propagator provided that it is massive. A simple free-particle
Yukawa-type massive propagator would be enough and the corrections due to the interaction would then be
manageable by perturbation theory.

A first attempt along these lines was reported in Ref.\cite{ptqcd} where, by a second order massive
expansion, the gluon and ghost propagators are evaluated and found in fair agreement with the lattice data.
The integrals were regularized by a simple cutoff that breaks the BRST symmetry and gives rise to several
drawbacks like quadratic divergences and the need of a mass counterterm. However, a fine tuning of the
mass parameter seems to cure the drawbacks yielding an optimized expansion that reproduces the lattice data.

In this paper, the difficulties of dealing with a cutoff are avoided by the use of a more robust
dimensional regularization scheme, yielding a more rigorous perturbative study of
pure $SU(N)$ Yang-Mills theory from first principles. While the original Lagrangian is not changed in any way,
the outcome is a one-loop analytical description that is infrared safe and in striking agreement with the
data of lattice simulations. Moreover the result can be improved by including higher-order terms and by use of
standard Renormalization Group (RG) techniques for reducing the effect of higher order terms.

A very interesting property of the massive expansion is the cancellation of all diverging mass terms without
including any spurious mass counterterm. Only wave function renormalization constants
are required and, in the minimal subtraction scheme, these constants are the same of the standard perturbative
expansion, thus ensuring that the correct UV behaviour is recovered.

The massive expansion is discussed in the Landau gauge in the present paper. The Landau
gauge is probably the optimal choice for the expansion, because of the transversality of the propagator
that makes the longitudinal polarization irrelevant. 
In the Landau gauge the problem decouples and a fully analytical result can be found for the propagators at one-loop.

While massive models have been studied before and found in good agreement with the 
data of lattice simulations\cite{tissier10,tissier11,tissier14},
the present calculation is very different because the Lagrangian is not modified, overall BRST symmetry is not
broken and no free parameters are added to the exact Yang-Mills theory, 
yielding a description that is based on first principles and can be improved order by order.
Thus, at variance with previous massive models, the present method would not give a mass to the photon.

The paper is organized as follows:
in Section II the massive expansion is developed for pure $SU(N)$ Yang-Mills theory in a generic
covariant gauge; in Section III the double expansion is set up in the Landau gauge; in Section IV
the explicit cancellation of the diverging mass terms is discussed in detail; in Section V explicit
analytical expressions are derived for the propagators at one-loop; in Section VI the one-loop propagators
and their scaling properties are compared with the available lattice data; in Section VII the running
coupling is evaluated and its sensitivity to the renormalization conditions is discussed, showing that
the approximation is under full control below $300$ MeV; finally, in Section VIII the main results are discussed.
Explicit analytical expressions for the propagators are given in the Appendix.

\section{Set up of the massive expansion in a generic gauge}

Let us consider pure Yang-Mills  $SU(N)$ gauge theory without
external fermions in a $d$-dimensional space. 
The Lagrangian can be written as
\BE
{\cal L}={\cal L}_{YM}+{\cal L}_{fix}+{\cal L}_{FP}
\EE
where ${\cal L}_{YM}$ is the Yang-Mills term
\BE
{\cal L}_{YM}=-\frac{1}{2} \Tr\left(  \hat F_{\mu\nu}\hat F^{\mu\nu}\right)
\EE
${\cal L}_{fix}$ is a gauge fixing term and ${\cal L}_{FP}$ is the ghost term
arising from the Faddev-Popov determinant.

In terms of the gauge fields, the tensor operator $\hat F_{\mu\nu}$ is 
\BE
\hat F_{\mu\nu}=\partial_\mu \hat A_\nu-\partial_\nu \hat A_\mu
-i g \left[\hat A_\mu, \hat A_\nu\right]
\EE
where
\BE
\hat A^\mu=\sum_{a} \hat X_a A_a^\mu
\EE
and the generators of $SU(N)$ satisfy the algebra
\BE
\left[ \hat X_a, \hat X_b\right]= i f_{abc} \hat X_c
\EE
with the structure constants normalized according to
\BE
f_{abc} f_{dbc}= N\delta_{ad}.
\label{ff}
\EE
If a generic covariant gauge-fixing term is chosen
\BE
{\cal L}_{fix}=-\frac{1}{\xi} \Tr\left[(\partial_\mu \hat A^\mu)(\partial_\nu \hat A^\nu)\right]
\EE
the total action can be written as $S_{tot}=S_0+S_I$ where the free-particle term is
\begin{align}
S_0&=\frac{1}{2}\int A_{a\mu}(x)\delta_{ab} {\Delta_0^{-1}}^{\mu\nu}(x,y) A_{b\nu}(y) {\rm d}^dx{\rm d}^dy \nn \\
&+\int \omega^\star_a(x) \delta_{ab}{{\cal G}_0^{-1}}(x,y) \omega_b (y) {\rm d}^dx{\rm d}^dy
\label{S0}
\end{align}
and the interaction is
\BE
S_I=\int{\rm d}^dx \left[ {\cal L}_{gh} + {\cal L}_3 +   {\cal L}_4\right].
\label{SI}
\EE
with the three local interaction terms that read
\begin{align}
{\cal L}_3&=-g  f_{abc} (\partial_\mu A_{a\nu}) A_b^\mu A_c^\nu\nn\\
{\cal L}_4&=-\frac{1}{4}g^2 f_{abc} f_{ade} A_{b\mu} A_{c\nu} A_d^\mu A_e^\nu\nn\\
{\cal L}_{gh}&=-g f_{abc} (\partial_\mu \omega^\star_a)\omega_b A_c^\mu.
\label{Lint}
\end{align}
In Eq.(\ref{S0}), $\Delta_0$ and ${\cal G}_0$ are the standard free-particle propagators for
gluons and ghosts and their Fourier transforms are
\begin{align}
{\Delta_0}^{\mu\nu} (p)&=\Delta_0(p)\left[t^{\mu\nu}(p)  
+\xi \ell^{\mu\nu}(p) \right]\nn\\
\Delta_0(p)&=\frac{1}{-p^2}, \qquad {{\cal G}_0} (p)=\frac{1}{p^2}.
\label{D0}
\end{align}
Here the transverse and longitudinal projectors are defined as
\BE
t_{\mu\nu} (p)=\eta_{\mu\nu}  - \frac{p_\mu p_\nu}{p^2};\quad
\ell_{\mu\nu} (p)=\frac{p_\mu p_\nu}{p^2}
\label{tl}
\EE
where $\eta_{\mu\nu}$ is the metric tensor. 

A shift of the pole in the gluon propagator can be introduced by an unconventional
splitting of the total action. Since we have the freedom of adding and subtracting the same
arbitrary term $\delta S$ to the total action
\begin{align}
S_0&\rightarrow S_0+\delta S\nn\\
S_I&\rightarrow S_I-\delta S
\label{shift}
\end{align}
we can take
\BE
\delta S= \frac{1}{2}\int A_{a\mu}(x)\>\delta_{ab}\> \delta\Gamma^{\mu\nu}(x,y)\>
A_{b\nu}(y) {\rm d}^dx{\rm d}^dy
\label{dS1}
\EE
where the vertex function $\delta\Gamma$ is a shift of the inverse propagator
\BE
\delta \Gamma^{\mu\nu}(x,y)=
\left[{\Delta_m^{-1}}^{\mu\nu}(x,y)- {\Delta_0^{-1}}^{\mu\nu}(x,y)\right]
\label{dG}
\EE
and ${\Delta_m}^{\mu\nu}$ is a massive free-particle propagator 
\begin{align}
{\Delta_m^{-1}}^{\mu\nu} (p)&={\Delta_m}(p)^{-1} t^{\mu\nu}(p)  
+\left[\frac{-p^2}{\xi}+{\cal A}(p)\right]\ell^{\mu\nu}(p)\nn\\
{\Delta_m}(p)^{-1}&=-p^2+{\cal M}(p)^2.
\label{Deltam}
\end{align}
Here the dynamical mass ${\cal M}(p)$ and the longitudinal function ${\cal A}(p)$ are left
totally arbitrary. While the total action cannot depend on them, just because $\delta S$ is added 
and subtracted again, any expansion in powers of the new shifted interaction $S_I\to S_I-\delta S$ is going 
to depend on the choice of $\delta S$ because of the approximation. 
Thus, it is the approximation 
that changes but we are not changing the content of the exact theory.
The shift $\delta S$ has two effects: the free-particle propagator ${\Delta_0}^{\mu\nu}$ is replaced by
the massive propagator ${\Delta_m}^{\mu\nu}$ in $S_0$; a counterterm $-\delta S$ is added to the interaction
$S_I$. 

From now on, let us drop all color indices in the diagonal matrices.
Inserting Eq.(\ref{D0}) and (\ref{Deltam}) in Eq.(\ref{dG}) the counterterm reads
\BE
\delta \Gamma^{\mu\nu} (p)={\cal M}(p)^2 t^{\mu\nu}(p) +{\cal A}(p)\ell^{\mu\nu}(p)
\label{dG2}
\EE
and must be added to the standard vertices arising from Eq.(\ref{Lint}).

The proper gluon polarization $\Pi$ and ghost self energy $\Sigma$ can be evaluated, order by order,
summing up Feynman graphs where ${\Delta_m}^{\mu\nu}$ is the zeroth order gluon propagator.
By Lorentz invariance we can write
\BE
\Pi^{\mu\nu}(p)=\Pi^T(p) t^{\mu\nu}(p)+\Pi^L(p) \ell^{\mu\nu}(p)
\label{pol}
\EE
and the dressed propagators are
\begin{align}
\Delta_{\mu\nu}(p)&=\Delta^T (p)t_{\mu\nu}(p)+\Delta^L (p)\ell^{\mu\nu}(p)\nn\\
{\cal G}^{-1}(p)&=p^2-\Sigma (p)
\end{align}
where the transverse and longitudinal parts read
\begin{align}
{\Delta^T}^{-1} (p)&=\left[-p^2+{\cal M}(p)^2-\Pi^T(p)\right]\nn\\
{\Delta^L}^{-1} (p)&=\left[\frac{-p^2}{\xi}+{\cal A}(p)-\Pi^L(p)\right].
\label{DTL}
\end{align}

At tree level, the polarization is just given by the counterterm $\delta \Gamma$ of Eq.(\ref{dG2}),
so that the tree-terms $\Pi^T_{tree}={\cal M}^2$, $\Pi^L_{tree}={\cal A}$ just cancel the shifts in the
dressed propagator $\Delta$ of Eq.(\ref{DTL}), giving back the standard free-particle propagator
of Eq.(\ref{D0}).
In fact, at tree-level, nothing really changes. 

Summing up all loops, the exact dressed propagator
can be written as
\begin{align}
{\Delta^T}(p)&=\left[-p^2-\Pi^T_{loops}(p)\right]^{-1}\nn\\
{\Delta^L}(p)&=\frac{\xi}{-p^2-\xi\Pi^L_{loops}(p)}.
\label{DTLloop}
\end{align}
As a consequence of gauge invariance, the exact longitudinal polarization $\Pi^L_{loops}$ must be zero
and the longitudinal part of the exact propagator must be equal to its tree-level value
$\Delta^L=-\xi/p^2$, just because the loop-terms cannot change it, as recently confirmed
by lattice simulations\cite{binosi15}.
Since $\Pi^L_{loops}$ and $\Pi^T_{loops}$ are evaluated by insertion of the modified propagator 
${\Delta_m}^{\mu\nu}$ in the loops, they can be considered as functionals of the arbitrary functions
${\cal M}$, ${\cal A}$. Thus, summing up all loops, the following constraints must hold for the
exact polarization functions:
\begin{align}
&\frac{\delta \Pi^T_{loops}}{\delta {\cal A}}=\frac{\delta \Pi^T_{loops}}{\delta {\cal M}}=0\nn\\
&\Pi^L_{loops}[{\cal A},{\cal M}]=0.
\label{constraint}
\end{align}

Expanding in powers of the total interaction, including the counterterm $\delta \Gamma$ among the vertices
and writing down the Feynman graphs, we can truncate
the expansion at a finite order yielding approximate functionals that may not satisfy the constraints
of Eq.(\ref{constraint}) exactly. For instance, the exact vanishing of the transverse polarization would be
lost unless BRST symmetry is mantained order by order. Actually, while the total Lagrangian has not been
changed and mantains its symmetry, the two parts $S_0$ and $S_I$ might be not BRST invariant because
of the arbitrary shift $\delta S$. Then the exact symmetry is lost in the expansion at any finite order
and the constraints are expected to hold only approximately unless all the graphs are summed up.
The outcome of the truncated expansion becomes sensitive to the choice of the functions ${\cal A}$, ${\cal M}$
thus suggesting the use of Eq.(\ref{constraint}) as variational stationary conditions.

Since we expect that the approximation should work better if the zeroth order propagator $\Delta_m^{\mu\nu}$
is a good approximation of the exact one, then  comparing Eq.(\ref{DTLloop}) and Eq.(\ref{Deltam}), 
a self-consistent method could be set up by requiring that 
\BE
{\cal M}(p)^2=\Pi^T_{loops}[{\cal M}]; \quad {\cal A}=\Pi^L_{loops}=0.
\label{selfcons}
\EE
Summing up all the loops these equations would be equivalent to SDE. Variational methods of this kind
have been investigated in several works\cite{gep2,varqed,varqcd,genself} and require the solution
of integral equations that can hardly be treated analytically.

In the Landau gauge the problem decouples and a fully analytical result can be found for the propagators. 
The Landau gauge is very special as the gluon propagator is transverse and does not depend on the choice of the
function ${\cal A}$. In the limit $\xi\to 0$ the longitudinal part $\Delta^L$ is exactly zero in
Eq.(\ref{DTLloop}) and is decoupled from the longitudinal polarization that becomes irrelevant for the
calculation of the propagator. In fact, in the same limit, the zeroth order propagator ${\Delta_m}^{\mu\nu}$ 
in Eq.(\ref{Deltam}) becomes transverse for any choice of ${\cal A}$ and the longitudinal part of
the counterterm $\delta \Gamma$ does not give any contribution in the loops when sandwiched by two transverse
propagators. Thus the only action of ${\cal A}$ is at tree level where it cancels itself in the gluon propagator
and disappears. Then, in the Landau gauge,
we can safely drop all longitudinal parts and set ${\cal A}=0$ without affecting the calculation. 
Because of the decoupling, the calculation of the longitudinal and of the transverse parts can be seen as 
two separate problems that may even require different orders of approximation. That simplifies things 
considerably, since a poor approximation for the longitudinal polarization would not affect the accuracy 
of the propagator. 
That also explains why reliable results for the propagator can be achieved even when the BRST symmetry is
broken\cite{tissier11} in the Landau gauge.
Moreover, since the total Lagrangian has not been modified, the overall BRST symmetry is unbroken and the
constraints in Eq.(\ref{constraint}) must be satisfied asymptotically. Thus, if required, a better approximation
for the longitudinal polarization can always be achieved in a separate calculation 
by adding more terms to the expansion.

\section{Double expansion in the Landau Gauge}

The Landau gauge is probably the optimal choice for the massive expansion, as discussed in the previous section.
In the limit $\xi\to 0$ the gluon propagators are transverse exactly and, having set ${\cal A}=0$, 
we can simplify the notation and drop the projectors $t^{\mu\nu}$ everywhere whenever each term is
transverse.
Moreover, we make the minimal assumption of taking the arbitrary function ${\cal M}$ equal
to a constant mass scale ${\cal M}=m$. In fact, variational calculations seem to suggest\cite{varqcd,genself}
that the actual form of the zeroth order propagator is not important provided that it is massive.
A constant mass simplifies the calculation and allows the use of dimensional regularization.

\begin{figure}[b] \label{fig:counterterm}
\centering
\includegraphics[width=0.09\textwidth,angle=-90]{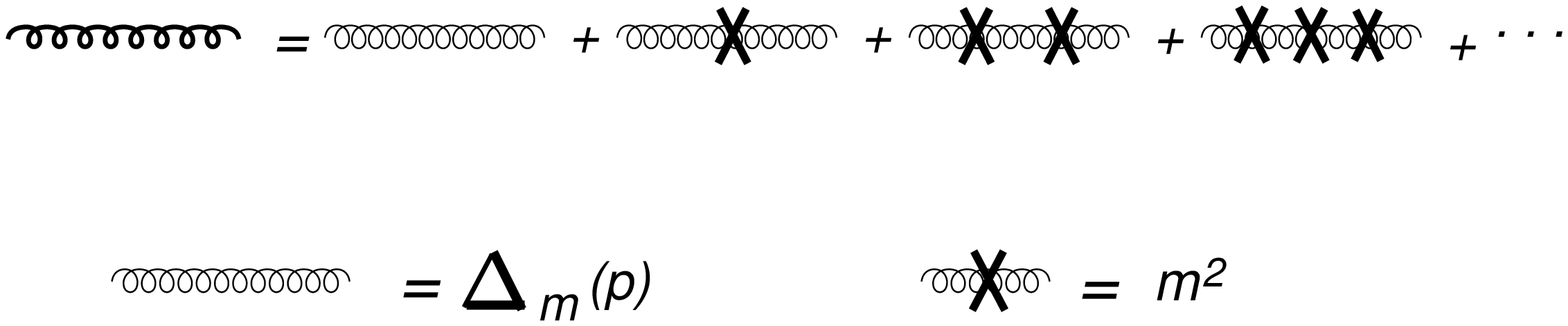}
\caption{Graphical illustration of Eq.(\ref{geometric}). The cross is the counterterm of Eq.(\ref{dGm})
that gives a factor $m^2$.}
\end{figure}

We can use the standard formalism of Feynman graphs with a massive zeroth
order propagator that reads
\BE
\Delta_m(p)=\left[-p^2+m^2\right]^{-1}
\label{Dm}
\EE
and a counterterm
\BE
\delta \Gamma =m^2
\label{dGm}
\EE
that must be added to the standard three-particle ghost-gluon and gluon-gluon vertices
of order ${\cal O} (g)$ and to the four-particle gluon-gluon vertex of order ${\cal O} (g^2)$
according to Eq.(\ref{Lint}). Thus, the total interaction is a mixture of terms that depend on 
the coupling strength $g$ and a counterterm that does not vanish in the limit $g\to 0$.
A perturbative expansion in powers of the total interaction would contain at any order different
powers of $g$ but the same number of vertices (including the counterterm among vertices) and 
we may define the order of a term as the number of vertices in the graph.
Of course, we could easily sum up some infinite set of graphs, like the chain graphs in Fig.1.
Formally, summing up all graphs with $n$ insertions
of the counterterm in the internal gluon lines, would cancel the pole shift and would give back the standard
perturbation theory 
\BE
\frac{1}{-p^2+m^2}\sum_{n=0}^{\infty}\left[ m^2 \frac{1}{-p^2+m^2}\right]^n=\frac{1}{-p^2}.
\label{geometric}
\EE

In fact, this is what we would get exactly at tree level, as discussed in the previous section.
That just says that the massive and the standard expansions are equivalent if we sum up all graphs.
On the other hand, at any finite order, the massive expansion is not 
equivalent to the standard perturbation theory, but the two expansions differ by an infinite class of graphs
that amounts to some non-perturbative content.  For instance, the massive zeroth order
propagator $\Delta_m$ cannot be obtained by the standard perturbation theory at any finite order because
of the gauge invariance of the theory that does not allow any shift of the pole.
We may reverse the argument and observe that, while in the UV the geometric expansion in Eq.(\ref{geometric})
is convergent and the two perturbation theories must give the same result, 
when $p^2\to m^2$ each single term in  Eq.(\ref{geometric}) diverges and the formal sum of infinite poles 
amounts to some non-perturbative content that makes the theories different at any finite order.
We can predict that the scale $m$ should be close to the Landau pole $\Lambda$ where the 
standard perturbation theory breaks down.

\begin{figure}[b] \label{fig:graphs}
\centering
\includegraphics[width=0.25\textwidth,angle=-90]{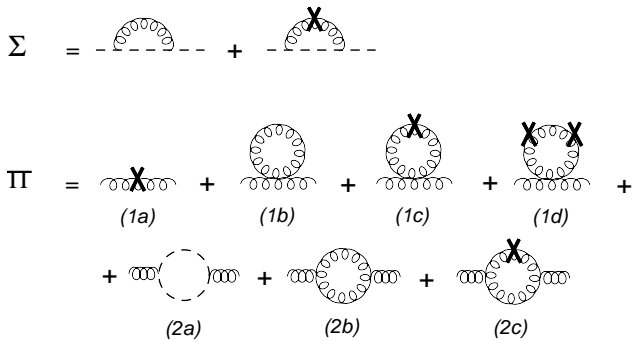}
\caption{Two-point graphs with no more than three vertices and no more than one loop. 
In the next sections, the ghost self energy and the gluon polarization  
are obtained by the sum of all the graphs in the figure.}
\end{figure}

Since we know that the gluon develops a dynamical mass in the infrared, we do not want to sum the chain
graphs in Fig.1 but prefer to truncate the power expansion at some finite order. An
expansion in powers of the total interaction $S_I$ is more efficient than the
standard expansion in powers of the coupling $g$. The counterterm $\delta \Gamma$ has the important
effect of reducing the weight of the total interaction since, in principle, if the zeroth order
propagator were exact, the total polarization would be exactly zero. For that reason, we define the order 
of a graph as
the number of vertices that are included, reflecting the power of $S_I$ rather than the number of
loops. Thus the tree-level graphs must be regarded as first order. As shown in Fig.2 the one-loop
tadpole $(1b)$ is first order while the gluon loop $(2b)$ is second order. Any insertion of the
counterterm $\delta \Gamma$ increases the order by one.

If the effective coupling is small, as it turns out to be according to non-perturbative calculations, 
not all the graphs have the same weight in the expansion. Since the number of loops is equal to the power 
of $g^2$ in a graph, two-loop graphs must be much smaller than one-loop and tree graphs.
We can consider a double expansion in powers of the total 
interaction and in powers of the coupling: we expand up to the $n$th order, retaining graphs with $n$ vertices
at most, and then neglect all graphs with more than $\ell$ loops. In Fig.2 the lower order graphs are
shown up to $n=3$ and $\ell=1$.

A very important feature of the double expansion is that there is no need to include mass counterterms 
for regularizing the divergences. All diverging mass terms cancel exactly in the expansion.
Thus we avoid to insert spurious mass parameters that were not in the original Lagrangian.
The cancellation can be easily explained by the following argument. Since we did not change the Lagrangian
and it was renormalizable (without mass counterterms because of BRST symmetry), all the diverging mass terms
must cancel if we sum up all graphs. In fact, no diverging mass term arises in the standard perturbative
expansion. The cancellation must be given by the sum of infinite graphs with counterterm insertions 
$\delta \Gamma$ in the loops that, according to Eq.(\ref{geometric}) and Fig.1, restore the pole
of the propagator and cancel the mass. However, if we inspect the graphs in Fig.2, we can easily see that 
any insertion of $\delta \Gamma$ in a loop reduces the degree of divergence of the graph so that they
become finite after a finite number of insertions.
Thus, if the divergences must cancel, they will cancel at a finite order of the expansion
provided that we retain more counterterm insertions than loops.
If $n$ is large enough, then all divergences in the mass terms are cancelled  by the counterterms in the loops. 
For instance, at one loop we only need $n=3$ as shown in Fig.2.

While in this paper we report the results for a  one-loop (third-order) approximation,
the extension to higher loops is straightforward and the regularization follows the same path
of standard perturbation theory, with all the divergences that can be cancelled by the usual
wave function renormalization constants.

\section{Cancellation of mass divergences in dimensional regularization}

The exact cancellation of the diverging mass-terms can be carried out explicitely in
dimensional regularization expanding in powers of $\epsilon=4-d$.

The insertion of one counterterm in a loop can be seen as the replacement 
\BE
\frac{1}{-p^2+m^2}\to\frac{1}{-p^2+m^2} m^2\frac{1}{-p^2+m^2}=-m^2\frac{\partial}{{\partial}m^2}\Delta_m
\label{deriv}
\EE
in the internal gluon line. If there are no other counterterm insertions in the same graph, 
then the dependence on $m^2$ must come from the massive propagators and a derivative of the whole 
$n$th-order $\ell$-loop graph gives the sum of all $(n+1)$-order $\ell$-loop graphs 
that can be written by a single insertion of $\delta\Gamma$ in any position.

In dimensional regularization, any diverging mass term that arises from a loop can be expressed
as a pole $c\> m^2/\epsilon$ where $c$ is a factor. Inserting this term in Eq.(\ref{deriv}), we see
that a counterterm in the loop gives a crossed-loop graph with the opposite diverging
term $-c\> m^2/\epsilon $. The argument also suggests a simple
way to evaluate the crossed-loop graphs by Eq.(\ref{deriv}).

At one-loop, we must truncate the expansion at the order $n=3$ for a full cancellation of
all the diverging mass terms. While higher-order terms could be included without introducing any
further divergence at one-loop, in this paper we explore the minimal approximation and sum up all
graphs up to $n=3$ as shown in Fig.2. 
It is not difficult to show that in the limit $p\to 0$ the gluon polarization is finite but not zero.
The existence of a finite limit $\Pi(0)\not=0$ is  crucial for the existence of a finite gluon
propagator in the infrared. 

First of all, let us evaluate the constant graphs at the order $n=3$.
At the lowest order ($n=1$, $\ell=0$) the counterterm $\delta \Gamma$ gives the 
constant graph $\Pi_{1a}=m^2$ that cancels the shift of the pole in the propagator. 
Exact integral expressions for the loop graphs have been reported by other authors in the Landau gauge.
In Ref.\cite{genself} all one-loop graphs are reported for any gauge, any space dimension and 
any choice of the zeroth order propagator. In Landau gauge and Euclidean space, 
the constant tadpole $\Pi_{1b}$ can be written as
\BE
\Pi_{1b}=-\frac{N g^2 (d-1)^2}{d} \int\kkd\frac{1}{k^2+m^2}.
\label{Ipi1b}
\EE
Expanding around $d=4$, in the $\overline{MS}$ scheme,
\BE
\Pi_{1b}=\frac{3}{4} \alpha\>m^2\>\left(\frac{2}{\epsilon}+\log\frac{\mu^2}{m^2}+\frac{1}{6}\right)
\EE
having hided the factor $N$ inside an effective coupling $\alpha$ defined as 
\BE
\alpha=\frac{3N}{4\pi} \alpha_s;\qquad \alpha_s=\frac{g^2}{4\pi}.
\EE

The crossed tadpole $\Pi_{1c}$ follows by a derivative according to Eq.(\ref{deriv})
\BE
\Pi_{1c}=-m^2\frac{{\partial}\Pi_{1b}}{{\partial} m^2}=
-\frac{3}{4} \alpha\>m^2\>\left(\frac{2}{\epsilon}+\log\frac{\mu^2}{m^2}-\frac{5}{6}\right).
\EE
As expected, the diverging terms cancel in the sum $\Pi_{1b}+\Pi_{1c}$. In fact, 
the double-crossed tadpole $\Pi_{1d}$ is finite and including its symmetry factor it reads
\BE
\Pi_{1d}= \frac{1}{2} m^4\frac{{\partial^2}\Pi_{1b}}{{\partial} (m^2)^2}=   -\frac{3}{8} \alpha\>m^2
\EE
so that the sum of the constant graphs is
\BE
\Pi_{1b}+\Pi_{1c}+\Pi_{1d}=\frac{3}{8} \alpha\>m^2.
\label{CG}
\EE 

While the ghost loop vanishes in the limit $p\to 0$, a finite $\Pi(0)\not=0$ can also
arise from the gluon loop $\Pi_{2b}$ that in the Landau gauge (in Euclidean space) 
can be written as\cite{genself}
\BE
\Pi_{2b} (p)=2Ng^2\int\kkd\frac{k_\perp^2{\cal F}(k,p)}{(k^2+m^2)[(k+p)^2+m^2]}
\label{Ipi2b}
\EE
where $k_\perp^2=[k^2-(k\cdot p)^2/p^2]$ and the kernel ${\cal F}$ can be decomposed as
\BE
{\cal F} (k,p)=\frac{k^2+p^2}{k^2}+\frac{p^2}{(k+p)^2}-\frac{p^2k_\perp^2}{(d-1)(k+p)^2 k^2}
\label{kernF}
\EE

The calculation of this graph is straightforward but
tedious. The integral can be evaluated analytically and the result is reported in the next section.
If we take the limit $p\to 0$ before integrating, we find that ${\cal F}\to 1$ and a mass term arises
\BE
\Pi_{2b} (0)=\frac{2Ng^2(d-1)}{d}\int\kkd\frac{k^2}{(k^2+m^2)^2}.
\label{Ipi2b0}
\EE
The integral is trivial and expanding around $d=4$ in the $\overline{MS}$ scheme we can write it as
\BE
\Pi_{2b}(0)=-\alpha m^2\left(\frac{2}{\epsilon}+\log\frac{\mu^2}{m^2}+{\rm const.}\right).
\EE
Adding the crossed loop $\Pi_{2c}$ with its symmetry factor, the divergences cancel
\BE
\Pi_{2b}(0)+\Pi_{2c}(0)=\left(1-m^2\frac{\partial}{\partial m^2}\right)\Pi_{2b}=-\alpha m^2
\EE
and adding the constant graphs in Eq.(\ref{CG}),
the one-loop dressed propagators can be written as
\begin{align}
\Delta(p)^{-1}&=-p^2+\frac{5}{8}\alpha m^2-\left[ \Pi(p)-\Pi(0)\right]\nn\\
{\cal G} (p)^{-1}&=p^2-\Sigma (p).
\label{dressed}
\end{align}
While the explicit calculation requires the evaluation of the gluon and ghost loop  
and of the ghost self-energy, 
we observe that a finite mass-term has survived in the cancellation, so that
the dressed propagator $\Delta(0)^{-1}=5\alpha m^2/8$ is finite and of order $\alpha$.

Actually, we checked that the full propagators in Eq.(\ref{dressed}), when renormalized, do
not depend on the precise value of the factor $5/8$ that arises by truncating
the expansion at the third order. A minor change of that coefficient is absorbed by a change
of the mass parameter and of the renormalization constants without affecting the final result.
That is an important feature since otherwise the whole calculation would depend on the somehow
arbitrary truncation of the expansion. In fact, while higher-order terms
would add very small corrections in the UV because of the factor $(-p^2+m^2)^{n+1}\sim(-p^2)^{n+1}$ in 
the denominators of Eq.(\ref{geometric}), in the limit $p\to 0$ the corrections might not be
negligible. In that limit we find a hierarchy in the significance of the crossed terms.
The most important effect arises at tree-level since the tree-graph $\Pi_{(1a)}$ in Fig.2 cancels
the entire shift of the pole in the propagator, as discussed in Section II. Thus, a finite
$\Pi(0)\not=0$ can only arise from loops and the massive expansion would not predict any mass
for the photon. At one-loop, a first insertion of the counterterm gives  diverging crossed
graphs that cancel the divergence of the loops entirely. Inclusion of those terms is crucial
for the renormalization of the theory. On the other hand, the insertion of $n$ counterterms in a loop,
with $n\ge 2$, gives finite terms that only add some fractions of $\alpha m^2$ to $\Pi (0)$. 
These terms decrease as $\sim 1/n^2$ and subtract each other, with a positive series of terms coming
from the tadpole graph and a negative series arising from the gluon loop.
Thus, the inclusion of higher order terms would only give a slight decrease of the coefficient of 
$\Pi(0)=-5\alpha m^2/8$ in Eq.(\ref{dressed}). 
That change is compensated by an increase of the mass parameter and by a change of the
renormalization constants, without making any real difference in the renormalized propagators.
In that sense, the minimal choice of a third order expansion has nothing special in itself and no dramatic effect is expected if higher-order terms are included.

\section{One-loop propagators}

The explicit evaluation of the propagators at one-loop and order $n=3$ requires the sum of the
gluon loop $\Pi_{2b}$, the crossed loop $\Pi_{2c}$ and the ghost loop $\Pi_{2a}$ for the gluon
propagator and the sum of the one-loop and crossed-loop self energy graphs for the ghost
propagator, as shown in Fig.2.

From now on, we switch to Euclidean space, expand the graphs around $d=4$ in the 
$\overline{MS}$ scheme and use the adimensional variable $s=p^2/m^2$.
The gluon loop $\Pi_{2b}$ is given by the integral in Eq.(\ref{Ipi2b}) that gives
a diverging part
\BE
\Pi^\epsilon_{2b}(p)=-\alpha \left( m^2-\frac{25}{36} p^2\right)
\left(\frac{2}{\epsilon}+\log\frac{\mu^2}{m^2}\right)
\label{2beps}
\EE
and a finite part
\begin{widetext}
\BE
\Pi^f_{2b}=\frac{\alpha m^2}{72}\left[\frac{2}{s}-135+\frac{226}{3}s+s^3\log s -s L_A(s)-s L_B(s)\right]
\label{2bf}
\EE
where $L_A$, $L_B$ are the logarithmic functions
\begin{align}
L_A(s)&=(s^2-20s+12)\left(\frac{4+s}{s}\right)^{3/2}
\log\left(\frac{\sqrt{4+s}-\sqrt{s}}{\sqrt{4+s}+\sqrt{s}}\right)\nn\\
L_B(s)&=\frac{2(1+s)^3}{s^3}(s^2-10s+1)\log(1+s).
\label{logs}
\end{align}
\end{widetext}

The ghost loop $\Pi_{2a}$ is a standard graph and in the Landau gauge it is given by the 
integral\cite{genself}
\BE
\Pi_{2a}(p)=-\frac{Ng^2}{(d-1)}
\int\kkd\>\frac{k_\perp^2}{k^2(p+k)^2}.
\label{Ipi2a}
\EE
The integral is straightforward and the diverging part is
\BE
\Pi^\epsilon_{2a} (p)=\frac{\alpha p^2}{36}
\left(\frac{2}{\epsilon}+\log \frac{\mu^2}{m^2}\right)
\label{2aeps}
\EE
while the finite part reads
\BE
\Pi^f_{2a} (p)=\frac{\alpha m^2}{36}
\left(2s-s\log s\right).
\label{2af}
\EE

The first of  self-energy graphs in Fig.2, the standard one-loop graph, 
is given by the integral\cite{genself}
\BE
\Sigma_1 (p)=-Ng^2 
\int\kkd\>\frac{ p^2 k_\perp^2}{k^2 (k-p)^2 (k^2+m^2)}
\label{Isi}
\EE
that yields a diverging term
\BE
\Sigma_1^\epsilon (p)=-\frac{\alpha p^2}{4}\left(\frac{2}{\epsilon}+\log \frac{\mu^2}{m^2}\right)
\label{seps}
\EE
and a finite part
\BE
\Sigma_1^f (p)=\frac{\alpha p^2}{12}\left[g(s)-5\right]
\label{sf}
\EE
where the function $g(s)$ is
\BE
g(s)=\frac{(1+s)^3}{s^2}\log(1+s)-s\log s-\frac{1}{s}.
\label{g}
\EE

If we do not add the crossed loops and take the sum of the finite parts, Eqs.(\ref{2bf}) and (\ref{2af}), then
we recover the finite part of the one-loop polarization function
\BE
\Pi_1^f (p)-\Pi_1^f (0)=-\frac{\alpha p^2}{72}\left[f(s)-\frac{238}{3}+\frac{111}{s}\right]
\label{pf}
\EE
where the function $f(s)$ is
\BE
f(s)=L_A(s)+L_B(s)+(2-s^2)\log s-2 s^{-2}.
\label{f}
\EE
We observe that in the limit $s\to 0$ the logarithmic functions have the limits $sL_A(s)\to -96$ and
$s(L_B(s)-2s^{-2})\to -15$, so that $s f(s)\to -111$.

The sum of the one-loop diverging parts, Eqs.(\ref{2aeps}) and (\ref{2beps}), 
yields the one-loop diverging term
\BE
\Pi_1^\epsilon (p)-\Pi_1^\epsilon (0)=\frac{13\alpha p^2}{18}
\left(\frac{2}{\epsilon}+\log \frac{\mu^2}{m^2}\right).
\label{peps}
\EE

Up to  irrelevant terms that depend on the renormalization scheme, the finite and diverging parts 
$\Sigma_1^f$, $\Pi_1^f$,$\Sigma_1^\epsilon$, $\Pi_1^\epsilon$,
as given by Eqs.(\ref{sf}), (\ref{pf}), (\ref{seps}) and (\ref{peps}), 
coincide with previous results\cite{tissier10,tissier11} for the one-loop functions with a massive propagator.

The crossed loops can be included very easily by a derivative with respect to $m^2$, as discussed in the
previous section below Eq.(\ref{deriv})
\begin{align}
\Sigma_{tot}&=\left(1-m^2\frac{\partial}{\partial m^2}\right)\Sigma_1=
\left(1+s\frac{\partial}{\partial s}\right)\Sigma_1\nn\\
\Pi_{tot}&=\Pi_{2a}+\Pi_{2b}+\Pi_{2c}=\left(1+s\frac{\partial}{\partial s}\right)\Pi_1
\label{crossed}
\end{align}
where we include all finite and diverging parts in the derivative.

The derivative of the diverging parts gives the finite terms $-\alpha p^2/4$ and
$13\alpha p^2/18$ that must be added to the finite parts of self-energy and polarization
function, respectively. Thus the diverging parts do not change and are given by the 
one-loop terms, Eqs.(\ref{seps}) and (\ref{peps}).

Performing the derivative of the finite parts we obtain
\begin{align}
\Sigma_{tot}^f (p)&=\frac{\alpha p^2}{12}\left[sg^\prime(s)+g(s)-8\right]\nn\\
\Pi_{tot}^f (p)-\Pi_{tot}^f (0)&=-\frac{\alpha p^2}{72} \left[sf^\prime(s)+f(s)-\frac{394}{3}\right]
\label{totf}
\end{align}
where $f^\prime$ and $g^\prime$ are the derivatives of $f$ and $g$, respectively.
The bare propagators follow by the insertion of finite and diverging parts in Eq.(\ref{dressed}).

The propagators can be made finite by the standard wave function renormalization. 
At one loop, the only residual
mass term is finite and of order $\alpha$, 
so that the divergences in Eq.(\ref{dressed}) are absorbed by the wave function
renormalization constants $Z_A$, $Z_\omega$.
In the $\overline{MS}$ scheme we find by Eqs.(\ref{seps}) and (\ref{peps})
\begin{align}
Z_A&=1+\frac{13\alpha}{9\epsilon}=1+\frac{13}{3}\frac{g^2 N}{16\pi^2}\frac{1}{\epsilon}\nn\\
Z_\omega&=1+\frac{\alpha}{2\epsilon}=1+\frac{3}{2}\frac{g^2 N}{16\pi^2}\frac{1}{\epsilon},
\label{z}
\end{align}
thus reproducing the same UV behaviour of the standard one-loop approximation.

It is useful to introduce the adimensional ghost and gluon dressing functions 
\BE
\chi (p)=-p^2{\cal G}(p);\quad J(p)=p^2 \Delta (p)
\label{dress1}
\EE
that, once renormalized by the constants $Z_A$, $Z_\omega$, are finite and read
\begin{align}
\chi (s)^{-1}&=1+\alpha\left[G(s)-\frac{2}{3}-\frac{1}{4}\log{\frac{\mu^2}{m^2}}\right]\nn\\
J (s)^{-1}&=1+\alpha\left[F(s)-\frac{394}{216}-\frac{13}{18}\log{\frac{\mu^2}{m^2}}\right]
\label{dress2}
\end{align}
where 
\begin{align}
F(s)&=\frac{5}{8s}+\frac{1}{72}\left[ s f^\prime (s)+f(s)\right]\nn\\
G(s)&=\frac{1}{12}\left[ s g^\prime (s)+g(s)\right].
\label{FG}
\end{align}
Explicit expressions for the {\it universal} functions $F(s)$, $G(s)$ are given in the Appendix.
Here we give the asymptotic behaviour. In the UV, for $s\gg 1$ we have
\BE
F(s)\approx\frac{17}{18} +\frac{13}{18}\log(s),\qquad G(s)\approx\frac{1}{3} +\frac{1}{4}\log(s)
\label{asympt}
\EE
while in the infrared, for $s\to 0$, we find that $G(s)$ tends to a constant and $F(s)\approx 5/(8s)$,
so that $\chi(0)$ is finite and $J(s)\approx 8s/(5\alpha)$, yielding
$\Delta (0)^{-1}=5\alpha m^2/8$ as expected from Eq.(\ref{dressed}).

We observe that in the UV, the asymptotic behaviour of Eq.(\ref{asympt}) is precisely what we need
for canceling the dependence on $m$ in the dressing functions. In fact, in the UV,  Eq.(\ref{dress2})
can be written as
\begin{align}
\chi (p)^{-1}&=\chi (\mu)^{-1}+\frac{\alpha}{4}\log\frac{p^2}{\mu^2}\nn\\
J (p)^{-1}&=J (\mu)^{-1}+\frac{13\>\alpha}{18}\log\frac{p^2}{\mu^2}
\label{dressUV}
\end{align}
which is the standard UV behaviour that we expected by inspection of the renormalization constants Eq.(\ref{z}).

The constants in Eq.(\ref{dress2}) have no direct physical meaning and depend on the special choice of 
renormalization constants in the $\overline{MS}$ scheme. We can subtract the dressing functions at a generic
point $s_0$ and, without fixing any special renormalization condition, we can write them in the more
general form
\begin{align}
\left[\alpha\> \chi (s)\right]^{-1}&=\left[\alpha\>\chi (s_0)\right]^{-1}+\left[G(s)-G(s_0)\right]\nn\\
\left[\alpha\> J (s)\right]^{-1}&=\left[\alpha\> J (s_0)\right]^{-1}+\left[F(s)-F(s_0)\right]
\label{dress3}
\end{align}
that extends the standard UV one-loop behaviour of the Eqs.(\ref{dressUV}), sharing with them
the same asymptotic behaviour for $s,s_0\gg 1$ according to Eq.(\ref{asympt}).
We observe that in general, we might not have the freedom of setting $J(s_0)=\chi(s_0)=1$
in Eq.(\ref{dress3}). Actually, $F(s)$ is not a monotonic function, it has a minimum and is
bounded from below, so that $J(s)^{-1}$ must also be bounded in Eq.(\ref{dress3}). 
Of course, that is just a limit of the 
one-loop approximation and the dressing functions can be renormalized at will by a different
choice of the renormalization constants. The point is that if the dressing functions are multiplied
by the arbitrary factor $Z=1+\alpha \delta Z$ then, at one-loop, that is equivalent to the subtraction
of $\alpha \delta Z$ on the right-hand sides of the Eqs.(\ref{dress2}). That only makes sense 
if $\delta Z$ is small and $Z\approx 1$. While in principle $Z$ can take any value, even much larger or
smaller than $1$, the one-loop subtraction can only compensate a small value of $\delta Z$. That is not
a problem in Eq.(\ref{dress3}) provided that we take account of any large renormalization factor by
direct multiplicative renormalization of $\chi(s_0)$ and $J(s_0)$. Then, if the energy $s$ is not too
far from the subtraction point $s_0$, the one-loop correction is small as it must be.

An important consequence is that by Eq.(\ref{dress3}) we can predict that at one-loop, up to an arbitrary
{\it multiplicative} renormalization constant, the inverse dressing functions are given by
the universal functions $F(s)$ and $G(s)$ up to an {\it additive} renormalization constant.
Such scaling property is satisfied quite
well by the lattice data, thus enforcing the idea that perturbation theory can provide important
insights on QCD in the infrared.

\section{Scaling properties on the lattice}

The predictive content of the theory can be tested by a direct comparison with the lattice
data. First of all, we would like to explore the scaling properties that emerge from
Eq.(\ref{dress3}) and that seem to be satisfied by the available lattice data for $SU(2)$ and
$SU(3)$. In fact, in Eq.(\ref{dress3}) any dependence on $\alpha$
is absorbed by the multiplicative renormalization constants of $\chi$ and $J$.
By such renormalization, the inverse dressing functions are entirely determined by
the universal functions $F(s)$ and $G(s)$ up to an additive constant.
In other words, by a special choice of the renormalization constants, all dressing functions can
be translated on top of the same curve by a vertical shift. In order to make that more explicit, 
we can write Eq.(\ref{dress3}) as
\begin{align}
\left[Z_G\>\chi (s)\right]^{-1}&=G(s)+G_0\nn\\
\left[Z_F \>J (s)\right]^{-1}&=F(s)+F_0
\label{dress4}
\end{align}
where $F_0$ and $G_0$ are a pair of constants depending on the subtraction point $s_0$, on the bare coupling
and on the normalization of the dressing functions $\chi(s_0)$, $J(s_0)$, while
$Z_G$, $Z_F$ are arbitrary renormalization constants that also absorb the dependence on $\alpha$.
While these equations predict a scaling property that is a stringent test for the one-loop approximation, 
the predictive content is remarkable: the  derivatives of the inverse dressing functions must be equal to
the derivatives of the universal functions $F(s)$, $G(s)$ up to an irrelevant multiplicative factor, 
while the additive constants $F_0$, $G_0$ emerge as unknown integration constants.

\begin{figure}[t] \label{fig:F}
\centering
\includegraphics[width=0.35\textwidth,angle=-90]{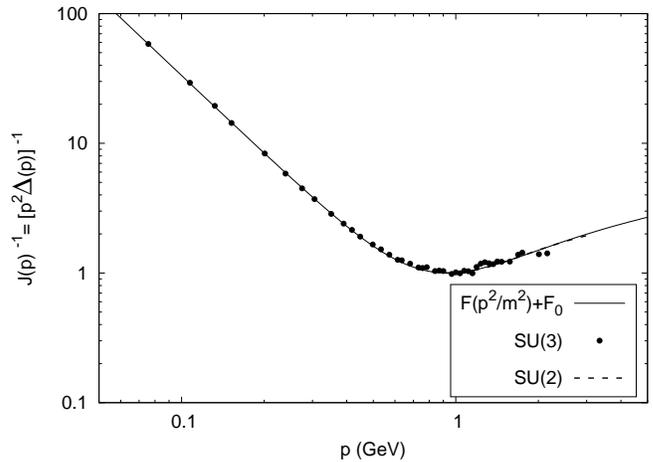}
\caption{The function $F(p^2/m^2)+F_0$  (solid line) is plotted together with the
lattice data for the 
inverse gluon dressing function $1/J(p)=p^2\Delta(p)$ renormalized by the factors in Table I according
to Eq.(\ref{dress4}).
The points are $SU(3)$ data extracted from a figure of Ref.\cite{bogolubsky} ($N=3$, $\beta=5.7$, $L=96$). 
The broken line is a fit of $SU(2)$ data by the empirical function of Ref.\cite{su2glu} ($\beta=2.40$, $L=42$),
only valid in the range $0.7-3$ GeV.
The energy scale is set by taking $m=0.73$ GeV for $N=3$ and $m=0.77$ GeV for $N=2$ (for the $SU(2)$ data the energy
is scaled by the ratio of the masses in order to superimpose the curves).}
\end{figure}

\begin{figure}[t] \label{fig:F2}
\centering
\includegraphics[width=0.35\textwidth,angle=-90]{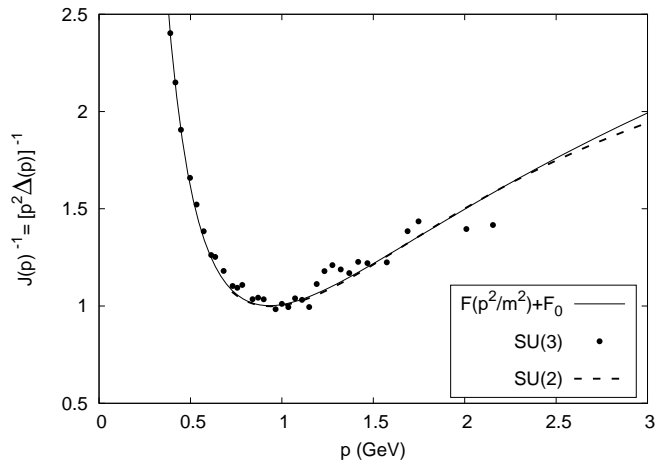}
\caption{Inverse gluon dressing function. An enlargement of the minimum area of Fig.3 is shown by 
a linear scale.}  
\end{figure}

\begin{figure}[t] \label{fig:G}
\centering
\includegraphics[width=0.35\textwidth,angle=-90]{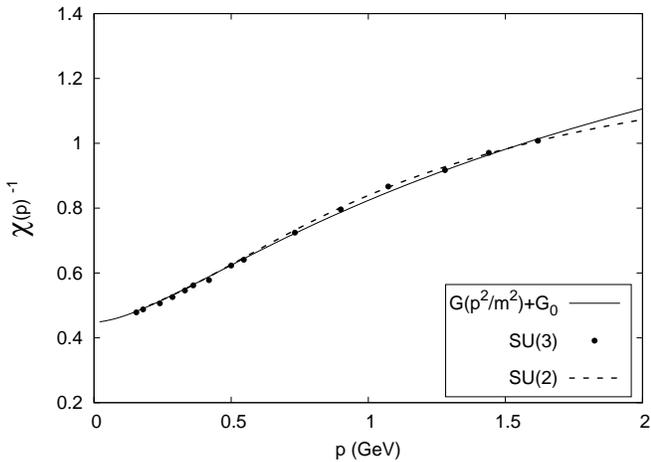}
\caption{The function $G(p^2/m^2)+G_0$  (solid line) is plotted together with the
lattice data for the 
inverse ghost dressing function $1/\chi(p)$ renormalized by the factors in Table I according
to Eq.(\ref{dress4}).
The points are $SU(3)$ data extracted from a figure of Ref.\cite{bogolubsky} ($N=3$, $\beta=5.7$, $L=80$). 
The broken line is a fit of $SU(2)$ data by the empirical function of Ref.\cite{su2gho} ($\beta=2.40$, $L=42$),
only valid in the range $0.2-3$ GeV.
The energy scale is set by taking $m=0.73$ GeV for $N=3$ and $m=0.77$ GeV for $N=2$ (for the $SU(2)$ data the energy
is scaled by the ratio of the masses in order to superimpose the curves).}
\end{figure}

The mass parameter $m$ provides the {\it natural} energy units that cannot be predicted by the theory and
can only be fixed by comparison with physical observables or lattice data. In fact, the total
Lagrangian does not contain any energy scale and, as for lattice calculations, the natural scale must
be regarded as a phenomenological quantity. However, once the mass $m$ is fixed, the original arbitrariness
of its choice is reflected in a spurious dependence on the subtraction point $s_0$ which is the only scale that
remains free in the theory. We expect that the residual dependence on $s_0$, which is implicit in the constants
$F_0$, $G_0$, should decrease if the approximation is improved by the inclusion of higher loops.

\begin{table}[ht]
\centering 
\begin{tabular}{c c c c c c} 
\hline\hline 
$N$ & \qquad $m$ (GeV)\quad & \quad $Z_G$ \quad & \quad $G_0$ \quad & \quad $Z_F$  
\quad & \quad $F_0$ \quad \\ [0.5ex] 
\hline 
2 & 0.77 & 0.888& 0.285 & 0.62 & -0.98 \\ 
3 & 0.73 & 0.637& 0.24  & 0.30 & -1.05 \\[1ex] 
\hline 
\end{tabular}
\label{table} 
\caption{Multiplicative and additive renormalization constants in Eq.(\ref{dress4}) and mass scales used
in the figures.} 
\end{table}

The function $F(s)+F_0$ is shown in Fig.3 together with the lattice data for the gluon inverse dressing function.
For $SU(3)$ the data points are extracted from a figure of Ref.\cite{bogolubsky} while for $SU(2)$ the
interpolation function of Ref.\cite{su2glu} is used, valid in the range 0.7-3.0 GeV. The data are scaled by
the renormalization constants in Table I and shown to collapse on the one-loop function $F(s)$ by a vertical
translation. Eq.(\ref{dress4}) is satisfied very well in the whole range of the lattice data. There is
a pronounced minimum that fixes the energy scale at $m=0.73$ GeV for $SU(3)$ and $m=0.77$ GeV for $SU(2)$.
In Fig.3 and Fig.4, the energy units of the data for $SU(2)$ have been scaled by the ratio of
the masses in order to superimpose them on the data for $SU(3)$. 
An enlargement of the area of  the minimum is shown in Fig.4 where the deviations
between the curves are amplified but found to be smaller than the fluctuations of the lattice data.

The function $G(s)+G_0$ is shown in Fig.5 together with the lattice data for the ghost inverse dressing function.
As in Fig.3, the lattice data for $SU(3)$ are extracted from a figure of Ref.\cite{bogolubsky} while the
data for $SU(2)$ are given by the interpolation function of Ref.\cite{su2gho}, valid in the range 0.2-3.5 GeV.
Again, the data are scaled by the renormalization constants in Table I and collapse on the one-loop function
$G(s)$ by a vertical translation. The energy units are the same of Fig.3 and Fig.4, i.e. the same values of $m$
are required for ghost and gluon dressing functions. We can see that the scaling properties 
predicted by Eq.(\ref{dress4}) are also satisfied very well by the ghost dressing function.

\begin{figure}[t] \label{fig:Delta}
\centering
\includegraphics[width=0.35\textwidth,angle=-90]{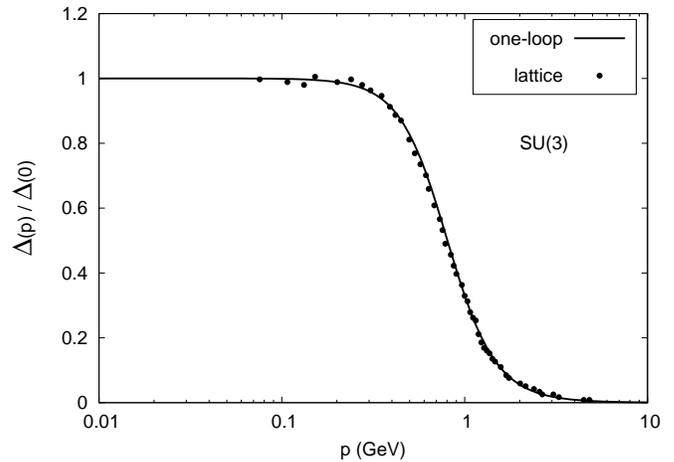}
\caption{The one-loop gluon propagator $\Delta (p)$ (line) 
is plotted together with the lattice data (points) extracted from
a figure of Ref.\cite{bogolubsky} ($N=3$, $g=1.02$, L=96) and   scaled
by the same renormalization constants of Table I. The energy scale is set by taking $m=0.73$ GeV.}
\end{figure}

\begin{figure}[t] \label{fig:chi}
\centering
\includegraphics[width=0.35\textwidth,angle=-90]{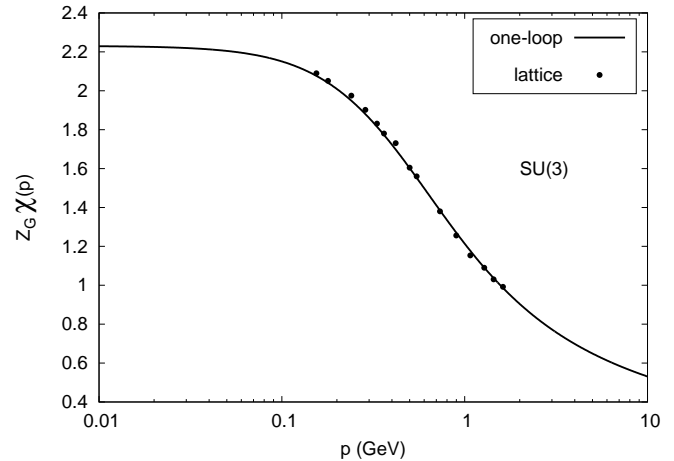}
\caption{The ghost dressing function $\chi (p)$  (line) is plotted together with the
Lattice data (points)  extracted from
a figure of Ref.\cite{bogolubsky} ($N=3$, $g=1.02$, L=80) and   scaled
by the same renormalization constants of Table I. The energy scale is set by taking $m=0.73$ GeV.}
\end{figure}

\begin{figure}[b] \label{fig:detail}
\centering
\includegraphics[width=0.35\textwidth,angle=-90]{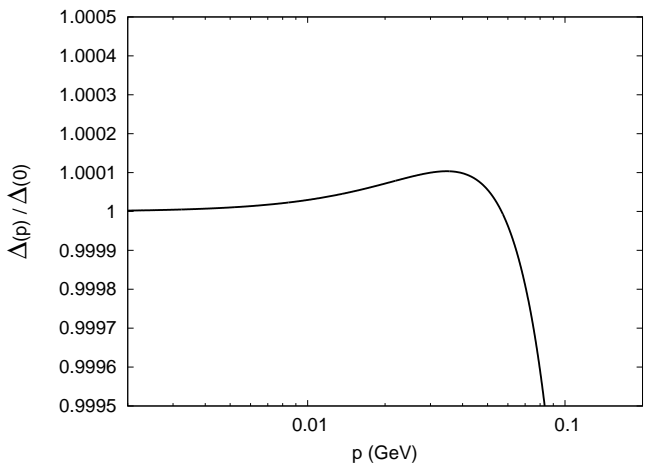}
\caption{Gluon propagator. Enlarged detail of Fig.6 deep in the infrared.}
\end{figure}

Overall, we find a very satisfactory description of the lattice data if the renormalization constants, the
multiplicative factors $Z_F$, $Z_G$ and the additive constants $F_0$, $G_0$ are fixed as in Table I.
While the multiplicative factors are not relevant anyway, we find a slight dependence on the additive constants
that cannot be compensated by a change of the factors, because of the one-loop approximation. 
Once the energy scale $m$ is fixed, no other free parameters are left besides the renormalization constants, 
so that the agreement with the lattice data is remarkable and really encouraging.

The gluon propagator and the ghost dressing function seem to show an even better accuracy than their inverse,
because of the scale.
For instance, for $SU(3)$ the gluon propagator and the ghost dressing function are reported in Fig.6 
and Fig.7, respectively, together with the lattice data of Ref.\cite{bogolubsky}. The renormalization constants
are set at the same values of Table I as discussed above.
We observe that the gluon propagator is not convex. Actually, it is not even a monotonic function of $p$, as shown
in Fig.8 where an enlargement of the deep infrared area is displayed in more detail. That property is usually assumed
to be a sign of confinement. A comparison of the gluon propagator with the lattice data of Ref.\cite{su2glu}
for $SU(2)$ is given in Fig.9.

\begin{figure}[t] \label{fig:gluonSU2}
\centering
\includegraphics[width=0.35\textwidth,angle=-90]{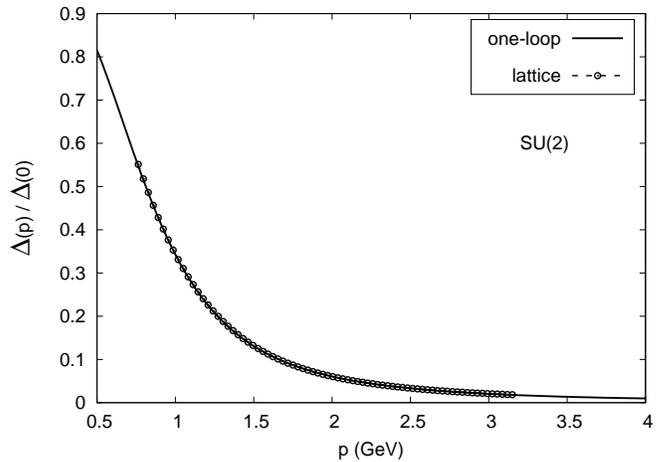}
\caption{The one-loop gluon propagator $\Delta (p)$ (solid line) 
is plotted together with the interpolation function of Ref.\cite{su2glu} (points) 
that fits the lattice data for
$SU(2)$ in the range $0.7-3$ GeV ($N=2$, $\beta=2.40$, $L=42$).
The renormalization constants of Table I are used. The energy scale is set by taking $m=0.77$ GeV.}
\end{figure}

\begin{figure}[b] \label{fig:alpha}
\centering
\includegraphics[width=0.35\textwidth,angle=-90]{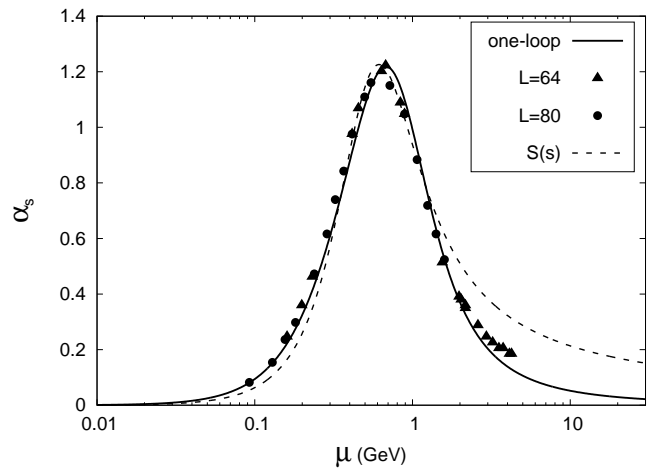}
\caption{The running coupling $\alpha_s(\mu)$ by Eq.(\ref{run}) (solid line) is
compared with the lattice data of Ref.\cite{bogolubsky} for $N=3$,
$\beta=5.7$, $L=64$ (triangles) and $L=80$ (circles). The constants $F_0$, $G_0$ and
the mass $m$ are set at the values of Table I for $N=3$. The coupling is
renormalized at the point $\mu=2$ GeV where we set $\alpha_s=0.37$. The broken line is
obtained by the function $S(s)$, according to Eq.(\ref{alps}), renormalized at the maximum
$\mu=0.67$ where $\alpha_s=1.21$.}
\end{figure}

\begin{figure}[b] \label{fig:alphauv}
\centering
\includegraphics[width=0.35\textwidth,angle=-90]{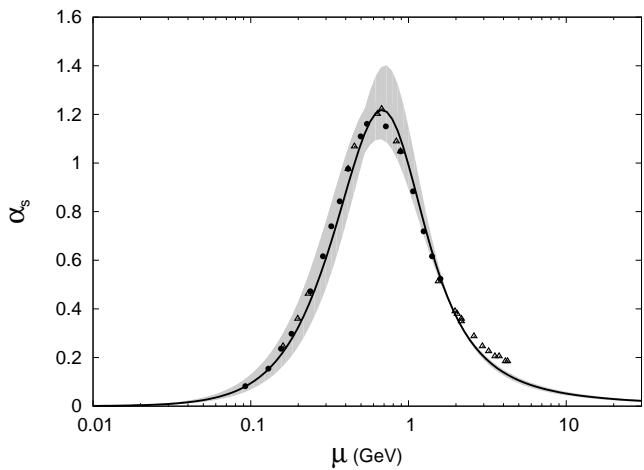}
\caption{ The filled grey pattern is the area spanned by the
coupling $\alpha_s(\mu)$ when the constants $F_0$, $G_0$ are changed
by $\pm 25\%$ with respect to the values in Table I for $N=3$. All couplings are
renormalized at $\mu=2$ GeV where $\alpha_s=0.37$.
The lattice data points and the solid line of Fig. 10 are superimposed for
comparison.}
\end{figure}

\begin{figure}[t] \label{fig:alphair}
\centering
\includegraphics[width=0.35\textwidth,angle=-90]{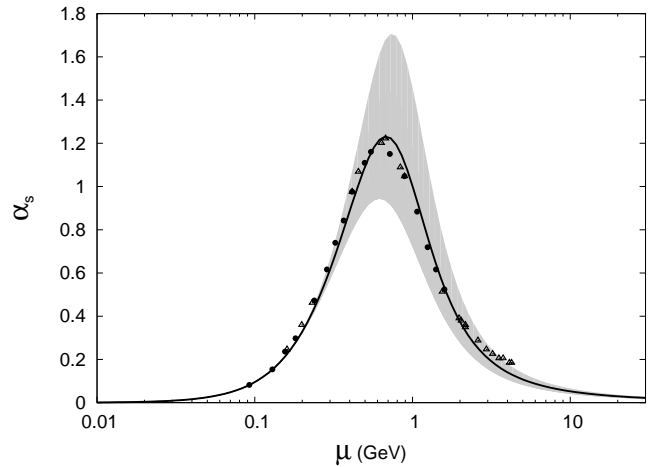}
\caption{The same as Fig. 11, but with all couplings renormalized in the infrared at $\mu=0.15$ GeV where
$\alpha_s=0.2$.}
\end{figure}

\begin{figure}[t] \label{fig:alphaM}
\centering
\includegraphics[width=0.35\textwidth,angle=-90]{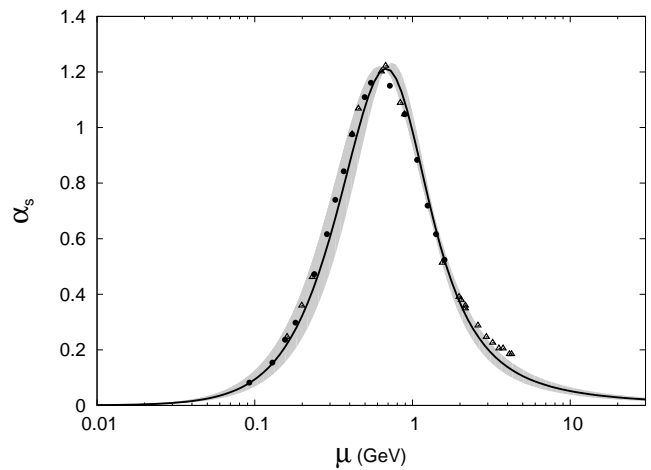}
\caption{The same as Fig. 11, but with all couplings renormalized at the maximum point $\mu=0.67$ GeV where
$\alpha_s=1.21$.}
\end{figure}

\section{Running coupling}

In the Landau gauge the ghost-gluon vertex 
is regular\cite{taylor} and the vertex renormalization constant can be set to one in a momentum-subtraction
scheme, so that a runnig coupling is usually defined by the RG invariant product of the dressing
functions
\BE
\alpha_s(\mu)=\alpha_s(\mu_0)\frac{ J(\mu) \chi(\mu)^2}{J(\mu_0) \chi(\mu_0)^2}.
\label{run0}
\EE
Having reproduced the dressing functions very well, we expect a very good agreement with the lattice for
the running coupling. Inserting Eq.(\ref{dress4}), the coupling reads
\BE
\alpha_s(\mu)=\alpha_s(\mu_0)\frac{ \left[F(\mu_0^2/m^2)+F_0\right] \left[G(\mu_0^2/m^2)+G_0\right]^2}
{\left[F(\mu^2/m^2)+F_0\right] \left[G(\mu^2/m^2)+G_0\right]^2}
\label{run}
\EE
and depends on the renormalization point $\mu=\mu_0$ where we set $\alpha_s(\mu_0)$ at a given
phenomenological value. We can renormalize the coupling at the point $\mu=2$ GeV where the
lattice data of Ref.\cite{bogolubsky}  give $\alpha_s=0.37$ for $SU(3)$. That is a good compromise
as the coupling is still quite small while the energy is not too large, so that we can still neglect the RG 
effects that become important in the UV limit\cite{tissier11}. We will refer to this point as the large energy
renormalization point. Using the values of Table I for $F_0$, $G_0$ and $m$ at $N=3$, the running coupling of
Eq.(\ref{run}) is displayed in Fig. 10, together with the lattice data of Ref.\cite{bogolubsky}.
The agreement is very good in the whole infrared range for $\mu<2.5$ GeV. In the UV, when $\mu> 2.5$ GeV,
we observe that Eq.(\ref{dress4}) starts to deviate from the lattice data. That is a known problem
that can be cured by a consistent running of the coupling 
in the one-loop calculation according to the RG equations, as shown in Ref.\cite{tissier11}.
On the other hand, in the infrared the agreement is impressive for a one-loop calculation.

It is instructive to explore how sensitive the result is to the choice of the additive renormalization constants
$F_0$, $G_0$, which are the only free parameters of the calculation. From a physical point of view, we would expect
that if the running coupling $\alpha_s(\mu)$ is the true effective coupling at the scale $\mu$, then the one-loop
approximation should be working very well deep in the infrared where $\alpha_s\to 0$. That would be very interesting
for future perturbative work. A test of the one-loop approximation comes from the sensitivity to changes of the
additive constants. If the approximation is under full control, then any small change of $F_0$ and $G_0$ 
should be compensated by the multiplicative renormalization constants, thus canceling in the normalized
ratio of Eq.(\ref{run}). In Fig.11, the grey pattern shows the area spanned by the running coupling $\alpha_s(\mu)$ of
Eq.(\ref{run}) when the additive renormalization constants $F_0$, $G_0$ are changed by $\pm25\%$ around the
values of Table I. Ignoring RG effects in the UV and comparing with the best running coupling of Fig.10,
which is also shown in the figure, we see that the deviations are very small in the UV and start growing up
when $\alpha_s\approx 0.6$. They increase until $\alpha_s$ reaches its maximum and then decrease getting smaller
and smaller in the infrared limit $\mu\to 0$. That enforces the idea that, deep in the infrared, the one-loop
approximation could be under full control. Moreover, the sensitivity to the additive constants seems to be
even smaller in the infrared if the renormalization point is taken at a very low energy.
In Fig. 12 the deviations are evaluated as before, by Eq.(\ref{run}), but renormalizing the coupling
at $\mu=0.15$ GeV where $\alpha_s=0.2$. We can see that the running coupling seems to be not
sensitive at all to the choice of the additive constants until $\alpha_s\approx 0.6$, and the approximation
seems to be under full control below $300$ MeV. In other words, regardless of the actual value of the
renormalization constants, all curves evaluated by Eq.(\ref{run}) collapse on the lattice data below $300$ MeV.
That is a remarkable feature as, by a proper choice of the renormalization point, the present one-loop
approximation provides a very accurate description of the running coupling below $300$ MeV (Fig. 12) or above 
$1.5$ GeV (Fig. 11), without adjusting any free parameter, from first principles.
In the range between $0.3$ and $1.3$ GeV, where $\alpha_s>0.6$, Eq.(\ref{run}) can be still tuned on the lattice data,
as shown in Fig. 10, but the increased sensitivity to the additive renormalization constants is a sign of the
limits of the one-loop approximation. However, since the calculation is from first principles, we expect that
the sensitivity to the additive constants should decrease when higher loops are included in the expansion.
 
From a technical point of view, Eq.(\ref{run}) provides a very good interpolation of the lattice data
and is not sensitive to the choice of the renormalization constants below $300$ MeV and above
$1.5$ GeV. Then, it could make sense to introduce a third fixed point by just renormalizing at the scale
where the deviations are larger and pinpoint $\alpha_s$ at its maximum. If we renormalize at the maximum
point $\mu=0.67$ GeV setting $\alpha_s=1.21$, the deviations are quite small over the whole range of energies,
as shown in Fig.13. That suggests that we can get rid somehow of the additive constants and write some
universal function for the running coupling, free of any parameter, albeit slightly approximate.

Let us pretend that we can set $\chi(s_0)=J(s_0)=1$ in Eq.(\ref{dress3}) and insert it in Eq.(\ref{run0}).
Then, neglecting higher powers of $\alpha$, the running coupling takes the simple shape
\BE
\alpha(s)=\frac{\alpha(s_0)}{1+\alpha(s_0)\left[S(s)-S(s_0)\right]}
\label{alps}
\EE
where $\alpha(p^2/m^2)=3N\alpha_s(p)/(4\pi)$ and the universal function $S(s)$ is defined as
\BE
S(s)=F(s)+2G(s)
\label{S}
\EE
and does not contain any free parameter.

In the UV, the running coupling $\alpha(s)$ incorporates the standard one-loop leading behaviour.
In fact, by Eq.(\ref{asympt}), for $s,s_0\gg 1$
\BE
\alpha[S(s)-S(s_0)]\approx 
\frac{11N\alpha_s}{12\pi}\log \left(\frac{p^2}{p_0^2}\right)
\label{alph1L}
\EE
which does not depend on the scale $m$. 
In the infrared, the function $S(s)$ replaces the standard log, yielding a finite running coupling 
without encountering any Landau pole.
In the limit $s\to 0$ the function diverges as $S(s)\sim (1/s)$ and the coupling goes 
to zero as a power $\alpha(s)\sim s$. 
A maximum is found at the point where ${\rm d}S(s)/{\rm d}s=0$, which occurs at $s_M= 1.044$. 
Of course, this point does not depend on any parameter and provides an independent way to fix the scale $m$ by
a comparison with the lattice. From the data of Ref.\cite{bogolubsky} in Fig.10 the maximum occurs at 
$p\approx0.6-0.7$ GeV yielding a scale $m\approx 0.6$ GeV, not too far from the values in Table I. 
Taking $m=0.6$ GeV and the maximum as renormalization point, namely $p=0.67$ GeV and $\alpha_s=1.21$ as in Fig. 13,
the plot of Eq.(\ref{alps}) is shown in Fig.10 as a broken line. 

The running coupling $\alpha(s)$ in Eq.(\ref{alps})
provides a nice qualitative description from first-principles, incorporates the standard leading UV behaviour
at one-loop and can be used  for extending the standard one-loop running coupling deep in the infrared.

\section{Discussion}

Let us summarize the main findings of the paper.

It has been shown that, from first principles, 
without changing the original Lagrangian, Yang-Mills theory can be studied
by a perturbative expansion by just taking a massive propagator as the expansion point.
Without the need to include spurious parameters or mass counterterms, the expansion can be renormalized
and all the divergences are canceled by the standard wave function renormalization of the fields.

At one-loop, the derivatives of the inverse propagators are determined, up to irrelevant multiplicative
factors, by the derivatives of the universal functions $F(s)$, $G(s)$, that do not depend on any
parameter. Thus, once a scale is fixed (the theory does not contain a scale that must come from the phenomenology),
the inverse dressing functions are determined up to an integration constant.
The relevant features of the dressing functions are contained in the universal functions $F$, $G$ regardless
of the specific value of the bare coupling and of $N$. That scaling property has been shown to be satisfied very well
by the lattice data, enforcing the idea that the infrared range of QCD can be studied by perturbation theory.

While the derivatives of the dressing functions are derived exactly, the propagators depend on the integration 
constants $F_0$, $G_0$. If the coupling is small and the one-loop approximation is under full control 
we would expect that a slight change of the additive constants could be compensated by a change of the 
irrelevant multiplicative factors. Actually, that only occurs in the UV and deep in the infrared where the
effective running coupling is small. In the range $0.5-1$ GeV, where the coupling reaches its maximum, the
propagators are sensitive to the choice of the additive constants. That seems to be a sign that higher
loops might be relevant when the effective coupling is larger. Thus, we expect that the sensitivity 
should decrease when higher loops are included in the calculation.
 
Even where the coupling $\alpha_s$ is not very small and two-loop corrections seem to be relevant, the one-loop calculation may acquire a variational meaning. The dependence on the renormalization constants is a consequence of an overall dependence on the ratio of the two energy scales: the mass parameter $m$ and the renormalization point $\mu$. Since the exact result should not depend on that ratio, the dependence is expected
to decrease when higher loops are included in the calculation. Thus a best choice for that ratio could be
obtained by some stationary condition on the observables, requiring that the sensitivity should be minimal in the
predicted phenomenology. However, there is no proof that a best choice of the renormalization constants does
exist, mimimizing two-loop corrections everywhere.
Thus, it is encouraging to know that, by tuning the additive constants, the one-loop calculation already 
provides an excellent description of the lattice data for the propagators and the running coupling. We conclude
that, while not anomalously small in general, two-loop corrections can be minimized by a best choice of
the constants.

Moreover, the sensitivity to the additive constants $F_0$, $G_0$ seems to be really negligible below $300$ MeV
and above $1.5$ GeV, namely when $\alpha_s<0.6$. In those ranges the running coupling collapses on the lattice
data without the need to tune any constant or parameter, from first principles and by a fully analytical 
description.

The existence of an energy range, deep in the infrared, where the one-loop approximation seems to be under
full control, could open the way for a more general analytical study of QCD below $\Lambda_{QCD}$ where
many interesting phenomena still suffer the lack of a full description from first principles.

\appendix
\section{Explicit functions $F$ and $G$}

The functions $F(x)$ and $G(x)$ are defined in Eq.(\ref{FG}) in terms of the functions $f(x)$, $g(x)$ and
their derivatives. Here we give the explicit expressions. The functions $f$, $g$ are given in Eqs.(\ref{f})
and (\ref{g}), respectively. They are encountered in the calculation of the standard one-loop polarization 
and self energy with a massive propagator and coincide with the result of other authors\cite{tissier10,tissier11}.
The derivatives are straightforward and have been checked by a software package. The result is
\begin{align}
F(x)&=\frac{5}{8x}+\frac{1}{72}\left[L_a+L_b+L_c+R_a+R_b+R_c\right]\nn\\
G(x)&=\frac{1}{12}\left[L_g+R_g\right]
\label{FGx}
\end{align}
where the logarithmic functions $L_x$ are
\begin{align}
L_a(x)&=\frac{3x^3-34x^2-28x-24}{x}\>\times\nn\\
&\times\sqrt{\frac{4+x}{x}}
\log\left(\frac{\sqrt{4+x}-\sqrt{x}}{\sqrt{4+x}+\sqrt{x}}\right)\nn\\
L_b(x)&=\frac{2(1+x)^2}{x^3}(3x^3-20x^2+11x-2)\log(1+x)\nn\\
L_c(x)&=(2-3x^2)\log(x)\nn\\
L_g(x)&=\frac{(1+x)^2(2x-1)}{x^2}\log(1+x)-2x\log(x)
\label{logsA}
\end{align}
and the rational parts $R_x$ are
\begin{align}
R_a(x)&=-\frac{4+x}{x}(x^2-20x+12)\nn\\
R_b(x)&=\frac{2(1+x)^2}{x^2}(x^2-10x+1)\nn\\
R_c(x)&=\frac{2}{x^2}+2-x^2\nn\\
R_g(x)&=\frac{1}{x}+2.
\label{rational}
\end{align}

\end{document}